\documentclass[eqsecnum,aps, twocolumn, superscriptaddress, 4pt ]{revtex4-1}
\usepackage{graphicx}
\usepackage{rotating}
\usepackage{epstopdf}
\usepackage{amsmath}
\usepackage{bm}
\usepackage{hyperref}

\begin{document}

\title {$E1$ PNC transition amplitudes of the hyperfine components for $^2S_{1/2}$ $-$ $^2D_{3/2}$ transitions of $^{137}$Ba$^{+}$ and $^{87}$Sr$^{+}$}
\author{\large Narendra Nath Dutta and \large Sonjoy Majumder \\
 \small {\it  Department of Physics, \\  Indian Institute of Technology-Kharagpur, \\
\small   Kharagpur-721302, India \\}}

\date{\today}

\begin{abstract}
\noindent In this paper, we have calculated parity nonconserving electric dipole transition amplitudes of the hyperfine components for the transitions between the ground and first excited states of $^{137}$Ba$^{+}$ and $^{87}$Sr$^{+}$ using sum-over-states technique. The results are presented to extract the constants associated with the nuclear spin dependent amplitudes from experimental measurements. The wavefunctions to calculate the most dominant part of the sums are constructed using highly correlated coupled-cluster theory based on the Dirac-Coulomb-Gaunt Hamiltonian.
\end{abstract}
\maketitle


The anapole moment (AM) is a parity violating electromagnetic moment of a nucleus \cite{flambaum1, flambaum2, dmitriev}. Calculations and measurements on parity non conservation (PNC)- induced electric dipole ($E1$) transitions in atomic systems is being considered as an excellent way to estimate this moment \cite{dzuba1, dzuba6}. It is the nuclear spin dependent (NSD) part of the PNC that provides the value of the AM of a nucleus \cite{dzuba1, flambaum1, flambaum2, dmitriev}. Whereas, the dominant nuclear spin independent (NSI) part of the PNC depends on the weak nuclear charge \cite{dzuba2}. Estimation of the anapole values of various nuclei is a promising tool to put constraints on the PNC meson-nucleon coupling constants \cite{dmitriev, liu, flambaum2}. The anapole constant of the $^{133}$Cs nucleus, which has a valence proton, was measured with near about 15$\%$ accuracy by Wood et al. \cite{wood, dmitriev, johnson}. However, this anapole value is found to be inconsistent with some results obtained through different nuclear many-body theories \cite{dmitriev} and also with the anapole value of $^{205}$Tl nucleus \cite{dmitriev, vetter}. Though, the latter was measured with very large uncertainty \cite{vetter}. Therefore, in this current situation, it is necessary to perform highly accurate PNC calculations and measurements on several species including systems having neutron as a valence nucleon to clear up this issue \cite{liu, dmitriev}. Also, the anapole values of this kind of systems can infer about the weak potential between the neutron and nuclear-core \cite{flambaum2, dmitriev}.

Isotopes of several singly ionized heavier ions like Ba$^{+}$, Ra$^{+}$, and Yb$^{+}$, having valence neutron in the nucleus, are considered as potential candidates of estimating anapole values through the PNC calculations and measurements \cite{dzuba1, sahoo1}. The experimental technique of reaching very accurate PNC measurement on the $6s$ $^2S_{1/2}$ $-$ $5d$ $^2D_{3/2}$ transition of Ba$^{+}$ was suggested by Fortson \cite{fortson}. This work is  going on at Seattle \cite{sherman1, sherman2, williams}. Theoretical calculations on a few  isotopes of Ba$^{+}$ in this regard were performed by Dzuba et al. \cite{dzuba1, dzuba4} and Sahoo et al. \cite{sahoo1, sahoo2}. The $E1$ NSD PNC amplitudes of the most recent calculations of Dzuba et al. \cite{dzuba1} differ by about 12$\%$ from the results of Sahoo et al. \cite{sahoo1}. The PNC results of Dzuba et al. are presented in the form of a ratio ($R$) of NSD to NSI amplitudes \cite{dzuba1}. The wavefunctions used in the PNC calculations by them reflect considerable discrepancies in the hyperfine $A$-values from precise experimental measurements for few relevant states \cite{dzuba1}. The hyperfine $A$ values are the most important tools to judge the accuracy of the wavefunctions of the states in contact to nuclear region where PNC interaction takes place. Nevertheless, as mentioned by them, such inaccuracies are canceled in the $R$ value associated with a hyperfine component if both the NSI and NSD amplitudes are calculated by similar approach considering all the leading contributions are accounted in a same way to them \cite{dzuba1, roberts2}.  

Though, the PNC effects are prominent in heavier systems, their theoretical accuracies are limited due to enhancement of the quantum electrodynamics (QED) correction \cite{dzuba3, dzuba5}, neutron skin effects \cite{brown} etc. Moreover, computational complexity for these systems to achieve desirable accuracy is more. Hence, it is reasonable to choose PNC candidates where both theory and experiment can keep conclusive accuracy. Recently, PNC calculations were carried out on relatively lighter system like $^{85, 87}$Rb \cite{dzuba3}. Following a similar trend, stable isotope like $^{87}$Sr$^{+}$ having a valence neutron in the nucleus may be considered as a good candidate for anapole estimation. PNC measurement on $5s$ $^2S_{1/2}$ $-$ $4d$ $^2D_{3/2}$ transition of Sr$^{+}$ was also proposed by Fortson using a similar technique as used for Ba$^{+}$ \cite{fortson}.

In the present work, we calculate the $E1$ PNC amplitudes of the hyperfine components (HC) for $6s$ $^2S_{1/2}$ $-$ $5d$ $^2D_{3/2}$  transition of $^{137}$Ba$^{+}$ using improved wavefunctions with respect to those used by Dzuba et al. \cite{dzuba1}. Also, we calculate these amplitudes of the HC for $5s$ $^2S_{1/2}$ $-$ $4d$ $^2D_{3/2}$ transition of $^{87}$Sr$^{+}$, for which no PNC data are available to the best of our knowledge. The results are presented with the aim of extracting the constants associated with the NSD PNC interactions using ongoing and future experiments \cite{dzuba6}. The anapole contributions to these constants can be extracted using a similar approach as discussed in detail in Ref. \cite{johnson, flambaum2}. In the present work, the sum-over-states technique is used where the main part or the dominating part \cite{blundell, pal, gopakumar1} of the sum is calculated with high accuracy using relativistic coupled-cluster (RCC) theory \cite{sur, lindgren, bishop} and experimental transition energies. The non-linear RCC theory with single, double and partial triple excitations (CCSD(T)) \cite{sur} is applied here to generate the $E1$ and weak matrix elements of this part in a correlation exhaustive way. Also, these matrix elements are generated here from the solutions of the Dirac-Coulomb-Gaunt (DCG) Hamiltonian to include the Gaunt correction \cite{dutta1}. Comparatively, less accurate method is used for the calculation of the rest part of the sum, where core polarization effect is included on top of the Dirac-Fock approximation based on the Dirac-Coulomb (DC) Hamiltonian. The PNC results of $^{137}$Ba$^{+}$ as obtained from the recent work of Dzuba et al. \cite{dzuba1} are compared with the present results as both the works are aiming the extraction of the anapole values using similar technique \cite{dzuba6}. 
   
The PNC interaction Hamiltonian due to both the NSI and NSD interactions is given by,  $H_{\text{PNC}}=H_{\text{NSI}}+H_{\text{NSD}}=\frac{G_F}{\sqrt{2}}\left( -\frac{Q_W}{2}\gamma_5+\frac{\kappa}{I}\boldsymbol{\alpha}.\boldsymbol{I}\right)\rho(r)$
\cite{dzuba1}. Here, G$_{F}$ is the Fermi constant of the weak interaction. $Q_W$ is the weak nuclear charge which is nearly equal to $-0.9877N+0.0716Z$, where $N$ and $Z$ are the number of neutrons and protons, respectively inside the nucleus \cite{dzuba1}. $\boldsymbol{\alpha}$ and $\gamma_5$ are the Dirac matrices. $\rho(r)$ is the normalized nuclear density distribution function, which is considered Fermi type here\cite{johnson}. 
$\kappa$ is a dimensionless constant which accounts the contributions from the anapole moment, electron-nucleus spin-dependent weak interaction and combined action of NSI PNC and hyperfine interaction \cite{johnson}. Using the sum-over-states technique, the spin independent and spin dependent $E1$ reduced matrix elements between the states $|J_f F_f\rangle$ and $|J_i F_i\rangle$ are derived as \cite{dzuba1, johnson, roberts},
\begin{eqnarray}
\langle J_fF_f||d_{\text{NSI}}||J_iF_i\rangle \nonumber
&=&(-1)^{I+F_i+J_f+1}\sqrt{[F_f][F_i]} \\ \nonumber
&\times & 
\left\{
\begin{array}{ccc}
  J_i & J_f & 1 \\
  F_f & F_i & I \\
\end{array}
\right\}\\ \nonumber & \times & \sum_n
\biggl[\frac{\langle J_f||d||J_n\rangle\langle J_n|
|H_{\text{NSI}}||J_i\rangle}{E_i-E_n} [J_i]^{-1/2}\\ \nonumber
&+& \frac{\langle J_f||
H_{\text{NSI}}||J_n\rangle\langle J_n||d||J_i\rangle}{E_f-E_n}[J_f]^{-1/2}\biggr],
\end{eqnarray} 
and 
\begin{eqnarray}
\langle J_f F_f||d_{\text{NSD}}||J_i F_i\rangle \nonumber
&=& \frac{\kappa}{I}\sqrt{I(I+1)(2I+1)[F_i][F_f]} \\ \nonumber
&\times & \sum_{n}\Biggl[ (-1)^{J_i-J_f+1} \\ \nonumber
&\times & \left\{
\begin{array}{ccc}
  F_f & F_i & 1 \\
  J_n & J_f & I \\
\end{array}
\right\}
\left\{
\begin{array}{ccc}
  I & I & 1 \\
  J_n & J_i & F_i \\
\end{array}
\right\} \\ \nonumber
&\times & \frac{\langle J_f||d||J_n\rangle\langle J_n|
|K||J_i\rangle}{E_i-E_n}\\ \nonumber
&+& (-1)^{F_i-F_f +1} \\ \nonumber
& \times & \left\{
\begin{array}{ccc}
  F_f & F_i & 1 \\
  J_i & J_n & I \\
\end{array}
\right\} 
\left\{
\begin{array}{ccc}
  I & I & 1 \\
  J_n & J_f & F_f \\
\end{array}
\right\}\\ \nonumber
& \times & \frac{\langle J_f||
K||J_n\rangle\langle J_n||d||J_i\rangle}{E_f-E_n}\Biggr].
\end{eqnarray}
where $[F]=2F+1$ and $[J]=2J+1$.
The single-particle reduced matrix elements of the operators $d$, $H_{\text{NSI}}$, and $K$ are given in Ref. \cite{johnson, roberts}. 

To judge the accuracy of the presently generated RCC wavefunctions based on the DCG Hamiltonian, we compare few results of the relevant properties of and among most important states for the PNC calculations with the highly accurate theoretical and most accurate experimental results. Table~\ref{Table:I} and Table~\ref{Table:II} depict the comparisons of these results for the hyperfine $A$ constants  and $E1$ transition amplitudes, respectively. These tables also contain the theoretical results calculated by Dzuba et al. \cite{dzuba1}. Here, the highly accurate theoretical results are estimated using the all-order single-double with partial triple (SDpT) excitations method by Safronova for both the systems Ba$^{+}$ \cite{safronovaba} and Sr$^{+}$ \cite{ safronovasr}. The experimental values are found from various earlier measurements where uncertainties were claimed most low \cite{nist, blatt, villemoes, silverans, sunaoshi, buchinger, pinnington, gallagher, davidson, kastberg, kurz}. Though, the $E1$ amplitude of $5d$ $^2D_{3/2}$ $-$ $6p$ $^2P_{1/2}$ transition is measured differently: 2.90(9) \cite{davidson}, 3.03(9) \cite{kastberg} and 3.14(8) \cite{sherman2} a.u. by different group with almost same precision. Also, in few cases, the experimental $E1$ transition amplitudes are estimated with large error bars. Therefore, we choice the SDpT results of the $E1$ transitions as standard to estimate the uncertainty in our PNC calculations as discussed later on. Nevertheless, the good agreements among our RCC results with the SDpT results and the experimental measurements (within the limit of uncertainty) as seen from these tables can ensure good quality of the RCC wave functions for all the states. Also, with respect to Ba$^{+}$, in case of Sr$^{+}$, one can find better agreement between the RCC and the SDpT values. Moreover, we have checked the Gaunt contributions to the $A$ constants of $6s$ $^2S_{1/2}$ and $5s$ $^2S_{1/2}$ states of $^{137}$Ba$^{+}$ and $^{87}$Sr$^{+}$, respectively. These values are +8.13 and +1.40 MHz, respectively and are consistent with the Breit contributions +8.33 and +1.39 MHz, respectively as obtained from Sushkov's analytic expression $\delta A=0.68ZA\alpha^2$ \cite{sushkov}. 

\begin{table}
\caption{Calculated hyperfine $A$ constants in MHz and their comparisons with the SDpT results of Safronova ($^{137}$Ba$^{+}$: \cite{safronovaba}, $^{87}$Sr$^{+}$: \cite{safronovasr}), theoretical results of Dzuba et al.  \cite{dzuba1}  and experimentally measured values (Exper.). The results of Dzuba et al. \cite{dzuba1} for $^{137}$Ba$^{+}$ are calculated by scaling their results for $^{135}$Ba$^{+}$ using the experimental ratio: $\frac{(A)_{^{137}\text{Ba}^+}}{(A)_{^{135}\text{Ba}^+}}$.}
\begin{tabular}{lcrrrr}
\hline\hline
Ions & States  &  Present & \cite{safronovaba, safronovasr} & \cite{dzuba1}  & Exper. \\
\hline
$^{137}$Ba$^{+}$& $6s$ $^2S_{1/2}$ & 4112.31 & 3997.39  & 4106 & 4018.87(0) \cite{blatt} \\
                & $6p$ $^2P_{1/2}$ &  731.13 &  733.98  &  747 & 743.7(0.3) \cite{villemoes} \\
                & $6p$ $^2P_{3/2}$ &  123.13 &  121.35  &  147 & 127.2(0.2) \cite{villemoes} \\
                & $5d$ $^2D_{3/2}$ &  194.18 &  191.53  &  180 & 189.73(0) \cite{silverans} \\
$^{87}$Sr$^{+}$ & $5s$ $^2S_{1/2}$ & 1008.35 &  997.85  &         & 1000.47(0) \cite{sunaoshi} \\
                & $5p$ $^2P_{1/2}$ &  175.70 &  177.33  &         &\\
                & $5p$ $^2P_{3/2}$ &   35.08 &   35.26  &         & 36.00(0.4) \cite{buchinger} \\
                & $4d$ $^2D_{3/2}$ &   46.62 &   46.70  &        & \\               
\hline\hline
\end{tabular}\\
\label{Table:I}
\end{table}

\begin{table}
\caption{Calculated $E1$ transition amplitudes in a.u. and their comparisons with the SDpT results of Safronova (Ba$^{+}$: \cite{safronovaba}, Sr$^{+}$: \cite{safronovasr}), theoretical results of Dzuba et al. \cite{dzuba1} and experimental values (Exper.). The Exper. of Sr$^{+}$ are calculated using the oscillator strengths given in the references and excitation energies from National Institute of Standards and Technology (NIST) \cite{nist}.}
\begin{tabular}{lcrrrr}
\hline\hline
Ions & Transition &  Present & \cite{safronovaba, safronovasr}  & \cite{dzuba1} & Exper. \\
\hline
Ba$^{+}$& $6s$ $^2S_{1/2}$ $-$ $6p$ $^2P_{1/2}$  & 3.3749 & 3.3710  & 3.32 & 3.36(4) \cite{davidson} \\
                & $6s$ $^2S_{1/2}$ $-$ $6p$ $^2P_{3/2}$  & 4.7586 & 4.7569  & 4.69 &  
  4.72(4) \cite{kurz} \\
                & $5d$ $^2D_{3/2}$ $-$ $6p$ $^2P_{1/2}$  & 3.0337 & 3.0957  & 3.06 &      
  3.03(9) \cite{kastberg} \\
                & $5d$ $^2D_{3/2}$ $-$ $6p$ $^2P_{3/2}$  & 1.3217 & 1.3532  & 1.34 &   
  1.36(4) \cite{kastberg}  \\                         
Sr$^{+}$& $5s$ $^2S_{1/2}$ $-$  $5p$ $^2P_{1/2}$  & 3.1059 & 3.0967   &  &  3.12 \cite{pinnington} \\
                & $5s$ $^2S_{1/2}$ $-$ $5p$ $^2P_{3/2}$  & 4.3891 & 4.3768 &   &  
4.40 \cite{pinnington} \\
                & $4d$ $^2D_{3/2}$ $-$ $5p$ $^2P_{1/2}$  & 3.0794 & 3.1193  &  & 3.47(32) \cite{gallagher}      
       \\
                & $4d$ $^2D_{3/2}$ $-$ $5p$ $^2P_{3/2}$  & 1.3669 & 1.3858  &   & 1.45(14) \cite{gallagher}   
       \\                                     
\hline\hline
\end{tabular}\\
\label{Table:II}
\end{table}

In Table~\ref{Table:III}, we present the $E1$ PNC transition amplitudes for the NSI interaction of $^{137}$Ba$^{+}$ and $^{87}$Sr$^{+}$ calculated by the present approach. In these calculations, the sums are considered for intermediate $np$ $^2P_{1/2}$ and $np$ $^2P_{3/2}$ states having values of $n$ from 2 to 25. The main parts of the sums contain  $n$=6, 7, and 8 for Ba$^{+}$ and $n$=5, 6 and 7 for Sr$^{+}$. These values of $n$ for the corresponding systems represent bound excited states at the Dirac-Fock (DF) level. The RCC theory is used here to construct the matrix elements of this part accurately. This theory can account the core correlation, core polarization and pair correlation contributions \cite{dutta3} to the matrix elements in an all order way \cite{sonjoy}. Even to increase the accuracy, the experimental transition energies are used from the NIST \cite{nist} at the denominators.
The Gaunt contributions to the pure DC values in the main parts of these amplitudes have been calculated around $-$0.4$\%$ and $-$0.3$\%$ for $^{137}$Ba$^{+}$ and $^{87}$Sr$^{+}$, respectively. On the top of these Gaunt corrected ab initio results, the replacement of our RCC energies by the NIST energies change the amplitudes by $+$1.2$\%$ to the former and by $+$1.1$\%$ to the latter. The main parts yield results 1.896 for $^{137}$Ba$^{+}$ and 0.260 for $^{87}$Sr$^{+}$ in the unit of $10^{-11}iea_0 Q_W/(-N)$.
The next contribution to the PNC amplitudes arises from the core or the auto-ionization states part of the sum \cite{pal}. This part takes the value of $n$ from 2 to 5 for Ba$^{+}$ and 2 to 4 for Sr$^{+}$. The remaining part or the tail part contributes little compare to the main and the core parts. In the core and tail sectors, core polarization (CP) effect is included in the matrix elements to provide sufficient accuracy in the final PNC amplitudes \cite{pal, blundell}. To include the CP effect in the core sector, we consider the second-order many-body perturbation theory (MBPT) diagram in Ref. \cite{kutzner} and replace this diagram by equivalent all-order diagram \cite{kutzner2} using Ref.\cite{sonjoy}. Similarly, in the highly excited valence sector or tail sector, we use a combination of all-order and second-order MBPT approach to incorporate the CP effect \cite{sonjoy, dutta3, sur}. With this treatment of the CP effect, for $^{137}$Ba$^{+}$, the core and tail contributions arise to 3.415 and 0.701, respectively at the DF+CP level which are 2.886 and 1.008, respectively at the DF level in the unit of $10^{-12}iea_0 Q_W/(-N)$. A large cancellation is seen to happen between the CP corrections to the core and the tail sectors. This similar kind of cancellation is seen to occur in the calculation of $E1$ NSI PNC amplitude for $7s$ $^2S_{1/2}$ $-$ $6d$ $^2D_{3/2}$ transition of $^{223}$Ra$^{+}$ by Pal et al \cite{pal}. They included the CP corrections in the core and the tail regions using random phase approximation (RPA) method. For $^{87}$Sr$^{+}$ also, the CP effect increases the core value from 3.732 to 4.364, but decreases the tail value from 0.863 to $-$0.166 in the unit of $10^{-13}iea_0 Q_W/(-N)$. 
   
In Table~\ref{Table:III}, we also compare the present $E1$ NSI PNC amplitude of $^{137}$Ba$^{+}$  with the other calculations obtained by the sum-over-states technique, but using different strategies. The results of Gopakumar et al. were evaluated by treating the main part at the RCC level, but the core and tail parts at the DF level \cite{gopakumar1}. Their DF orbitals are based on the hybridization of the Gaussian type orbital (GTO) bases and numerical bases \cite{gopakumar1}. Whereas, we use analytical GTO bases only to construct these orbitals. Theoretically, for more accurate treatment, the CP effect should be included at these less contributing parts, which is performed in the present approach. The work of Dzuba et al. used correlation potential method to generate the Bruckner type orbitals and included CP effect (using RPA method) in the relevant matrix elements for the PNC calculations including the hyperfine constants \cite{dzuba6, dzuba4, dzuba1}. As seen from Table~\ref{Table:I}, their method produces the hyperfine $A$-constants of the $6p$ $^2P_{3/2}$ and $5d$ $^2D_{3/2}$ states with a considerably large discrepancy from the experimental measurements and the SDpT values. Whereas, for the $6s$ $^2S_{1/2}$ and $6p$ $^2P_{1/2}$ states, the $A$-values obtained from all the three theories and the experiments are in a reasonably good agreement. Therefore, our correlation exhaustive RCC wavefunctions are more accurate on the average with respect to the wavefunctions of Dzuba et al. in and very near to the nuclear region. Also, both the calculations of Gopakumar et al. and Dzuba et al. did not consider the Gaunt corrections which is considered in the present approach. Nevertheless, our $E1$ NSI PNC amplitude agrees well with both the other results. This can be the consequence of cancellations between the various contributions to the sum.

\begin{table}
\caption{Calculated $E1$ NSI PNC amplitudes for the $6s$ $^2S_{1/2}$ $-$ $5d$ $^2D_{3/2}$ transition of $^{137}$Ba$^{+}$ and $5s$ $^2S_{1/2}$ $-$ $4d$ $^2D_{3/2}$ transition of $^{87}$Sr$^{+}$ in the unit of $10^{-11}iea_0 Q_W/(-N)$ . The present result of $^{137}$Ba$^{+}$ is compared with the corresponding results of Dzuba et al. \cite{dzuba4, dzuba1} and Gopakumar et al.  \cite{gopakumar1}.}
\begin{tabular}{lcrr}
\hline\hline
Ions & Present &   \cite{dzuba4, dzuba1}  & \cite{gopakumar1} \\
\hline
$^{137}$Ba$^{+}$ & 2.308  &  2.34 & 2.35 \\
$^{87}$Sr$^{+}$ & 0.302  &       &   \\
\hline\hline
\end{tabular}\\
\label{Table:III}
\end{table}

The results of Table~\ref{Table:V} is the major focus in the present work. These results are presented in the form of $\langle J_fF_f||d_{\text{NSI}}||J_iF_i \rangle[1+R\kappa]$, where $R=\frac{\langle J_fF_f||d_{\text{NSD}}||J_iF_i \rangle}{\kappa \langle J_fF_f||d_{\text{NSI}}||J_iF_i \rangle}$ \cite{dzuba1}. The NSD PNC amplitudes are calculated in the identical strategy that is adopted to calculate the NSI PNC amplitudes as explained earlier. To the former amplitudes, the main, core and the tail parts contribute about 80$\%$, 16.5$\%$ and 4.0$\%$, respectively in case of Ba$^{+}$ and about 82$\%$, 17.5$\%$ to 18.5$\%$ and $-$0.3$\%$ to 0.7$\%$, respectively in case of Sr$^{+}$. The results calculated by the present technique are compared with the results of Dzuba et al. for $^{137}$Ba$^{+}$ \cite{dzuba1}. In their paper, the values are presented in the z-component matrix element forms of hyperfine states \cite{dzuba1}. However, in the present comparison, we keep their results in reduced matrix element forms. Also, we invert the signs of their all NSI and NSD amplitudes to make these consistent with our sign conventions.   Using the DC Hamiltonian, we have found that the magnitudes of the $R$ values are changed by about $-$9.5$\%$ to $-$11.0$\%$ for $^{137}$Ba$^{+}$ and by about $-$17$\%$ to $-$24$\%$ for $^{87}$Sr$^{+}$ from the pure ab initio DF results to the correlation corrected results (RCC for the main sectors and DF+CP for the remaining sectors). These correlation corrections to the ratios are almost determined from the main parts of the sums. Therefore, exhaustiveness in correlation to the main parts is desirable to maintain excellent accuracy in the ratios. From these correlation corrected results, the $R$ values have been found to change by about +0.5$\%$ to +1$\%$ and +1.5$\%$ to +2$\%$ for $^{137}$Ba$^{+}$ and $^{87}$Sr$^{+}$, respectively to the final results as presented in the table. 

\begin{table}
\caption{Calculated $E1$ PNC amplitudes (reduced matrix elements) for the $|6s$ $^2S_{1/2}, F_i \rangle$ $-$ $|5d$ $^2D_{3/2}, F_f \rangle$ transition of $^{137}$Ba$^{+}$ and $|5s$ $^2S_{1/2}, F_i \rangle$ $-$ $|4d$ $^2D_{3/2}, F_f \rangle$ transition of $^{87}$Sr$^{+}$ in 10$^{-11}$ a.u. The results of $^{137}$Ba$^{+}$ are compared with the corresponding results of Dzuba et al. \cite{dzuba1}.}
\begin{tabular}{lcrrr}
\hline\hline
Ions & $F_f$ & $F_i$ & Present &  \cite{dzuba1}  \\
\hline
$^{137}$Ba$^{+}$ & 3 & 2 & $-$7.0166(1$-$0.0233$\kappa$) & $-$7.15(1$-$0.0239(2)$\kappa$)\\
                 & 2 & 2 &  4.1932(1$-$0.0231$\kappa$) &  4.27(1$-$0.022(1)$\kappa$)\\
                 & 2 & 1 & $-$4.1932(1+0.0386$\kappa$) & $-$4.27(1+0.038(1)$\kappa$)\\
                 & 1 & 2 & $-$1.8753(1$-$0.0229$\kappa$) & $-$1.93(1$-$0.021(2)$\kappa$)\\
                 & 1 & 1 &  4.1932(1+0.0387$\kappa$) &  4.29(1+0.0392(4)$\kappa$)\\
                 & 0 & 1 & $-$2.6520(1+0.0388$\kappa$) & $-$2.70(1+0.0398(3)$\kappa$)\\
$^{87}$Sr$^{+}$  & 6 & 5 & $-$1.2436(1$-$0.0335$\kappa$) & \\
                 & 5 & 5 &    0.8861(1$-$0.0351$\kappa$) & \\
                 & 5 & 4 & $-$0.7235(1+0.0433$\kappa$)   & \\
                 & 4 & 5 & $-$0.5343(1$-$0.0364$\kappa$) & \\
                 & 4 & 4 &   0.8861(1+0.0420$\kappa$)   & \\
                 & 3 & 4 & $-$0.9126(1+0.0409$\kappa$) & \\
\hline\hline
\end{tabular}\\
\label{Table:V}
\end{table}

Both the $E1$ NSI and NSD PNC amplitudes of $^{137}$Ba$^{+}$ and $^{87}$Sr$^{+}$ are calculated within theoretical uncertainty of about 3$\%$. These uncertainties are calculated using standard procedure \cite{dzuba1, pal} of replacing the $E1$ amplitudes obtained from the RCC theory by the $E1$ amplitudes calculated from the SDpT approximation and scaling a PNC amplitude $\langle f|H_{PNC}|i\rangle$ by the factor $\frac{\sqrt{(A_f \times A_i)_{\text{SDpT}}}}{\sqrt{(A_f \times A_i)_{\text{RCC}}}}$ and $\frac{\sqrt{(A_f \times A_i)_{\text{Exper.}}}}{\sqrt{(A_f \times A_i)_{\text{RCC}}}}$  in the main part \cite{pal}. Also, a rough approximation from the QED, neutron skin effects and more complete calculations in the core sectors are considered here. Nevertheless, the $R$ values for both the ions are calculated within theoretical uncertainty of about 0.5$\%$ considering the scaling of the matrix elements as mentioned above. Therefore, the ratio of two different precise PNC measurements corresponding to two different HC and it's comparison with the present theoretical value can lead to a very accurate interpretation of $\kappa$ \cite{dzuba6}. 
 
The PNC amplitudes of $^{137}$Ba$^{+}$ and $^{87}$Sr$^{+}$ have been calculated for the purpose of extracting the constants associated with the NSD PNC interactions with high accuracy from the ongoing experiment for the former ion and the proposed experiment for the latter ion. The amplitudes of $^{87}$Sr$^{+}$ are calculated for the first time. 
 
We want to acknowledge Prof. A. D. K. Singh, PRL, Ahmedabad, India for his  encouragement towards this work. We are thankful to Mr. S. Chattopadhaya, PRL, Ahmedabad, India for his valuable suggestions. We would like to recognize the
support from Council of Scientific and Industrial Research (CSIR), India to provide funding for our research.


\end{document}